\documentstyle[12pt] {article}
\begin{document}
\textheight 23 cm
\baselineskip 18 pt plus2pt
\topmargin -30.0 mm
{\hfill  TECHNION--PH-95-18}

\vspace{6.0cm}
\centerline{\Large Remarks on the formation of black holes in}
\centerline{\Large non-symmetric gravity}
\vspace{1.cm}
\centerline{Lior M. Burko and Amos Ori, Department of Physics}
\centerline{Technion -- Israel Institute of Technology, 32000 Haifa, Israel}
\vspace{2.5cm}
\centerline{Abstract}
In a recent paper,  we discussed the
formation of black holes in non-symmetric gravity.
That paper was then criticized by Cornish and
Moffat. In the present paper, we address the arguments raised by
Cornish and Moffat. In summary, we do not see any reason to doubt the
validity of our former conclusions.
\nopagebreak

\vspace{1.cm}
PACS numbers:  04.50+h  \hspace{0.3cm} 04.70.Bw
\newpage
\textheight 22 cm

Non-symmetric Gravitation Theory (NGT) is a generalized theory of
gravitation, which incorporates General Relativity (GR) as a limiting
case  \cite{moffat - review}. Recently, Cornish and Moffat (CM)
\cite{cornish and moffat} conjectured, based on a class of
exact {\it static} spherically-symmetric solutions, that NGT was free of
black holes. Later, Burko and Ori (BO) \cite{burko and ori} found that
there was no analogue in NGT to the Birkhoff theorem of GR \cite{clayton},
and realized that dynamics might have important consequences. In order
to get some insight into the dynamical content of NGT, BO studied in Ref.
\cite{burko and ori} the evolution of small, linearized, anti-symmetric
field over a symmetric GR background. The behavior of the linearized
skew perturbation was found to be perfectly regular in the entire
domain of dependence, and in particular at (and beyond) the event
horizon. Based on this result, BO concluded that if, in a situation of
gravitational collapse, the initial skew perturbation is sufficiently
small, then a black hole is likely to form, just like in GR.

More recently, the analysis of Ref. \cite{burko and ori} was criticized
by CM \cite{cornish and moffat 2}. In brief, the objection of CM is based
on the observation that NGT lacks a Principle of Equivalence, and that this
feature of NGT is not well captured by the linear approximation. CM
thus argue that the linear analysis is misleading, and that
the outcome of a
gravitational collapse in NGT is more likely to be a compact object
which is not a black hole.

In the present paper we briefly address the objection of CM. In brief,
we find this objection to be unjustified, and we do not see any reason
to doubt the validity of our former conclusions in Ref. \cite{burko and
ori}.

The main arguments raised in Ref. \cite{cornish and moffat 2} are the
following:

1)  The linear approximation to NGT is not valid, because it fails to
capture an important aspect of the fully non-linear NGT -- the
absence of an Equivalence Principle.

2)   The absence of the Equivalence Principle suggests that the perturbative
approach will break at the horizon: With no Equivalence Principle,
the evolving perturbations
are allowed to ``feel'' the horizon, by ``feeling'' the non-symmetric
field there.

3)   Consider a static star surrounded by a static
spherically-symmetric NGT field. The exterior field is described by the static
spherically-symmetric NGT solution -- the Wyman solution. On top of
this background, consider some additional small NGT perturbations.
Assume now that the star undergoes gravitational collapse.
CM argue that the final state cannot be a black hole, because of the
following reason: Since the above background is linearly stable, then,
upon evolution, the small initial deviation from the Wyman solution
must remain small. Therefore, the final state must be given by small
perturbations on top of the Wyman solution. Such a final state cannot
be a black hole (because the Wyman solution is not perturbatively
close to any black-hole solution).

Here is our reply to the above arguments:

1)   The absence of an Equivalence Principle in NGT is certainly an
interesting issue. But this issue has nothing to do with the validity of
the linear approach. For, in our analysis we did not make any use of the
Equivalence Principle, or any assumption about its existence or
inexistence in the theory.

Needless to say, the validity of a linear analysis has nothing to do
with the issue of whether the system in question admits an Equivalence
Principle or not. (The linear approach is used in countless number of systems
in physics, and in most of these cases, this principle is simply
irrelevant.)\footnotemark\footnotetext{
Perhaps the idea behind the above argument by CM is that the usage of
the linear approach should be limited to situations where the
background, or the linear field, admits all the qualitative aspects of
the non-linear system. This is certainly not the case: In many systems
where small perturbations were successfully analyzed by the linear
approach, there are important, or even crucial, non-linear phenomena
which are absent at the linear level. As an example, consider sound
waves in a perfect fluid
\cite{landau and lifshitz}: It is the non-linear aspects which are
responsible for the formation of shock waves, as well as turbulence and
other chaotic phenomena -- and yet the linear treatment is perfectly valid
(in a compact neighborhood)
if the initial perturbations are small. (A similar example is the
linearized Einstein theory.) Another example: In a vibrating elastic body
or string, coupling to higher harmonics occurs at the non-linear level
only, and yet the linear approach is valid for the description of small
vibrations.}

In addition, let us note that the most fundamental manifestation of
the break-down of the Equivalence Principle occurs already at the
linear level: The impossibility to find a local reference frame in which
the metric tensor is locally Minkowski, and the existence of nontrivial
metric-tensor scalars like  $g^{[\alpha \beta ]}g_{[\alpha \beta ]}$, are
faced already at the linear level. (As we have just pointed out,
however, this issue is absolutely irrelevant to the validity of the linear
approach!)

2)   We agree that (in a somewhat vague sense) the absence of the
Equivalence Principle allows the evolving perturbations to ``feel" the
non-symmetric field. But this has nothing special to do with the
horizon: The antisymmetric field has a non-zero value not only at $r=2m$,
but also at greater (or smaller)
$r$ values. Therefore, the absence of the Equivalence
Principle in the theory does not alter the following simple fact: The
event horizon is an absolutely regular location at the background, which
has no local significance\footnotemark \footnotetext
{We also point out that the results of Ref. \cite{burko and ori} are not a
consequence of the pre-existing black hole (and event horizon). As it
turns out, even if there is no black hole in the initial moment
({\it i.e.}, in the
gravitational collapse of matter or gravitational radiation), the
subsequent formation of
a black hole is to be anticipated \cite{burko and ori}.}.

3)   This argument of CM is based on the assumption that the initial
state is, in some global sense, ``close to Wyman", {\it i.e.},
that it can be
described by a small perturbation over Wyman. This assumption is
obviously wrong: If we denote the initial radius of the star
by $r_s$, then the
initial field is close to Wyman at $r>r_{s}$, but is {\it very }
different from
it at $r<r_s$. Thus, while in the Wyman solution the antisymmetric
field is typically very strong at $r=2m$, in the initial state considered here
it is presumably small there. In addition, the Wyman solution is vacuum
and static, and the initial state considered here is neither of them (at
$r<r_s$).\footnotemark \footnotetext{
This strong deviation of the initial state from Wyman considered here
has nothing to do with the extra small perturbations added on
top of the background; It is the {\it background} itself which is, in
overall, very different from Wyman.} Therefore, the expectation that
this configuration will evolve like Wyman with small perturbations has
nothing to base on.

Let us examine this issue more carefully: Let us denote the section
$r\ge r_s$ of the initial hypersurface by $S$.
Then, since the initial data are
close to Wyman on $S$, the standard stability arguments indeed suggest
that the evolving field will be similar to Wyman throughout
$D^+(S)$ -- but not elsewhere
(here it is important to recall that there is no Birkhoff theorem in NGT).

The region near $r=2m$ is not included in $D^+(S)$, and is not even
close to it. Therefore, there is no reason to expect that the evolving
field near $r=2m$ will by any means be similar to Wyman.

In fact, this last argument by CM can easily be reversed: If in
the initial configuration described in argument (3)
the initial skew field is
sufficiently small\footnotemark \footnotetext{
This would be the case if the parameter $s$ of the exterior Wyman
solution is sufficiently small.}
(which we assume), then the initial data can be viewed as a small NGT
perturbation on top of a GR background. The same line of reasoning
suggested by CM now implies that the evolving configuration will be
just GR plus small skew perturbation -- which means that a black hole
will form! Note that, unlike the original argument (3), this reversed
argument is indeed valid, because the initial data are close to a GR
configuration in the {\it entire} initial slice
(whereas, in contrast, they are
close to Wyman at $r>r_{s}$ only).

In summary, we find no reason to doubt the validity of the linear
approach of Ref. \cite{burko and ori}.
Therefore, although this linear analysis does not
provide a strict mathematical proof for anything (a linear analysis
almost never does), it
should be regarded as firm and trustable -- just like any other situation
in physics in which linear analysis is used to study the behavior of
small perturbations. We therefore insist that our conclusion in Ref.
\cite{burko and ori} is justified: In the situation of a
gravitational collapse in NGT, if the initial skew perturbation is
sufficiently small, one should expect a black hole to form -- just like in
GR.

\end{document}